\documentclass[aps,pra,reprint,showpacs]{revtex4-1}
\usepackage{color}
\usepackage[dvipdfmx]{graphicx}
\usepackage{amsmath,amssymb}
\usepackage{bm}

\begin{document}


\title{Metastable spin textures and Nambu-Goldstone modes of a ferromagnetic spin-1 Bose-Einstein condensate confined in a ring trap}


\author{Masaya Kunimi}
\affiliation{Department of Engineering Science, University of Electro-Communications, Tokyo 182-8585, Japan}
\email{E-mail: kunimi@hs.pc.uec.ac.jp}


\date{\today}

\begin{abstract}
We investigate the metastability of a ferromagnetic spin-1 Bose-Einstein condensate confined in a quasi-one-dimensional rotating ring trap by solving the spin-1 Gross-Pitaevskii equation. We find analytical solutions that exhibit spin textures. By performing linear stability analysis, it is shown that the solutions can become metastable states. We also find that the number of Nambu-Goldstone modes changes at a certain rotation velocity without changing the continuous symmetry of the order parameter.

\end{abstract}

\maketitle

\section{Introduction}\label{sec:Introduction}


Owing to recent developments in the experimental techniques associated with cold atomic gases, Bose-Einstein condensates (BECs) confined in multiply connected geometries have been realized \cite{Ryu2007,Ramanathan2011,Moulder2012,Beattie2013,Neely2013,Ryu2013,Wright2013,Wright2013_full,Ryu2014,Eckel2014,Jendrzejewski2014,Eckel2014_2,Corman2014}. Such systems are suitable for investigating the fundamental properties of superfluidity. In fact, many interesting features of superfluidity have already been observed, such as persistent current \cite{Ryu2007,Ramanathan2011,Moulder2012,Beattie2013}, phase slip and vortex nucleations \cite{Wright2013,Wright2013_full}, hysteresis \cite{Eckel2014}, and a current-phase relationship \cite{Eckel2014_2}.

Although the above experiments, except Ref.~\cite{Beattie2013}, were performed using scalar BECs, BECs with internal degrees of freedom (spinor BECs \cite{Kawaguchi2012,Stamper-Kurn2013}) confined in simply connected geometries have also been realized experimentally \cite{Stamper-Kurn1998,Stenger1998}. In spinor BECs, there exist various topological defects \cite{Leonhardt2000,Khawaja2001,Leanhardt2003,Ruostekoski2003,Kawaguchi2008,Kobayashi2009,Leslie2009,Wright2009,Choi2012,Choi2012_2,Ray2014}, spin textures \cite{Sadler2006,Vengalattore2008,Cherng2008,Lamacraft2008,Vengalattore2010}, and Nambu-Goldstone modes (NGMs) \cite{Ohmi1998,Ho1998,Murata2007,Uchino2010,Symes2014,Marti2014} due to their spontaneous symmetry breaking.

The previous theoretical works regarding multi-component systems (two-component Bose gases \cite{Smyrnakis2009,Bargi2010,Anoshkin2013,Smyrnakis2013,Wu2013} and spinor BECs \cite{Makela2013,Yakimenko2013}) in a ring trap concern the stability of persistent currents. Other important properties, including metastability under external rotation, have also been thoroughly investigated theoretically  \cite{Mueller2002,Kanamoto2009} and experimentally \cite{Wright2013,Wright2013_full,Eckel2014} for scalar BECs. However, a complete understanding of the metastability of spinor BECs in multiply connected systems under external rotation has not yet been established.

 In this paper, we investigate the properties of spin-1 BECs in a rotating ring trap within the framework of the mean-field approach. We present analytical solutions of the spin-1 Gross-Pitaevskii equation (GPE) \cite{Ohmi1998,Ho1998} under a twisted periodic boundary condition. This solution exhibits spin textures. By performing linear stability analysis, we show that these solutions can become metastable states. Furthermore, we determine the NGMs and find the change in the number of NGMs for a given rotational velocity without changing the continuous symmetry of the order parameter. This change in the number of NGMs is called a type-I$-$type-II transition \cite{Takahashi2014a}. 

\section{Model}

We consider $N$ spin-1 bosons confined in a rotating ring trap. Within the mean-field approximation, the system can be described by a three-component order parameter (a condensate wave function) $\bm{\Psi}(\bm{r}, t)\equiv [\Psi_1(\bm{r}, t), \Psi_0(\bm{r}, t), \Psi_{-1}(\bm{r}, t)]^{\rm T}$, where ${\rm T}$ denotes the transpose. For simplicity, we treat the system as a quasi-one-dimensional torus. We assume that the spatial dependence of the order parameter is given by $\bm{\Psi}(\bm{r}, t)\equiv \bm{\Psi}(x, t)/\sqrt{S}$, where $S$ is the cross section of the torus and $x$ represents the coordinate \cite{Note_angle}. The energy functional of the system is given by
\begin{align}
\hspace{-0.5em}E=\hspace{-0.3em}\int^{L}_{0}\hspace{-0.3em}dx\hspace{-0.3em}\left[\hspace{-0.1em}\frac{\hbar^2}{2M}\hspace{-0.1em}\sum_m\left|\frac{\partial  \Psi_m(x,t)}{\partial x}\right|^2\hspace{-0.6em}+\hspace{-0.2em}\frac{c_0}{2}n(x, t)^2\hspace{-0.2em}+\hspace{-0.1em}\frac{c_1}{2}\bm{F}(x, t)^2\hspace{-0.1em}\right]\hspace{-0.3em},\label{eq:Model_definition_of_Hamiltonian}
\end{align}
where $M$ is the mass of the boson; $m$ denotes the magnetic sublevels and can take the values of $1, 0$, and $-1$; $L$ is the length of the torus; $n(x, t)\equiv \sum_m|\Psi_m(x, t)|^2$ is the particle density; $\bm{F}(x, t)\equiv \sum_{m, n}\Psi_m^{\ast}(x,t)(\bm{f})_{m n}\Psi_n(x,t)$ is the magnetization density vector; $(f_{\nu})_{m n} (\nu=x, y, z)$ are the spin-1 matrices; $c_0\equiv 4\pi\hbar^2(a_0+2a_2)/3M S$ and $c_1\equiv 4\pi\hbar^2(a_2-a_0)/3MS$ are the spin-independent and spin-dependent interaction strengths, respectively; and $a_0$ and $a_2$ are the $s$-wave scattering lengths of the spin-0 and -2 channels, respectively.

The time-dependent GPE for spin-1 bosons \cite{Ohmi1998,Ho1998} is given by the functional derivatives $i\hbar\partial\Psi_m(x, t)/\partial t=\delta E/\delta\Psi_m^{\ast}(x, t)$,
\begin{align}
&i\hbar\frac{\partial}{\partial t}\Psi_{\pm 1}(x, t)=\mathcal{L}_{\pm 1}\Psi_{\pm 1}(x, t) +\frac{c_1}{\sqrt{2}}F_{\mp}(x, t)\Psi_0(x,t),\label{eq:Model_GP_equation_pm_1}\\
&i\hbar\frac{\partial}{\partial t}\Psi_0(x, t)=\mathcal{L}_0\Psi_0(x, t)\nonumber \\
&\hspace{3.0em}+\frac{c_1}{\sqrt{2}}\left[F_+(x,t)\Psi_1(x,t)+F_-(x,t)\Psi_{-1}(x,t)\right],\label{eq:Model_GP_equation_0}\\
&\mathcal{L}_m\equiv -\frac{\hbar^2}{2M}\frac{\partial^2}{\partial x^2}+c_0n(x, t)+mc_1F_z(x, t),\label{eq:Model_operator_L}
\end{align}
where $F_{\pm }(x,t)\equiv F_x(x,t)\pm iF_y(x,t)$. The effects of the rotation are described by imposing the twisted periodic boundary condition \cite{Lieb2002_2,Kunimi2014_1}
\begin{eqnarray}
\Psi_m(x+L,t)=\Psi_m(x,t)e^{iMvL/\hbar},\label{eq:Model_twisted_periodic_boundary_condition}
\end{eqnarray}
where $-v$ is the rotational velocity of the ring. This corresponds to the boundary condition in the rotating frame \cite{Kanamoto2009}. We note that the twisted boundary condition is invariant under the transformation $v\to v+l v_0$, where $v_0\equiv 2\pi\hbar/M L$ and $l\in \mathbb{Z}$.

The Bogoliubov equation for spin-1 bosons \cite{Kawaguchi2012} can derived by the linearizing the GPE around the stationary solution $\Psi_m(x, t)=e^{-i\mu t/\hbar}\Psi_m(x)$, where $\mu$ is the chemical potential of the system, which is determined by the total number of particles:
\begin{eqnarray}
N=\int^{L}_{0}dx \sum_m|\Psi_m(x)|^2.\label{eq:Model_total_particle_number_condition}
\end{eqnarray}
Substituting
\begin{align}
\Psi_m(x, t)&=e^{-i\mu t/\hbar}\left[\Psi_m(x)+u_m(x)e^{-i\epsilon t/\hbar}\right.\nonumber\\
&\hspace{12.0em}\left.-v^{\ast}_m(x)e^{i\epsilon^{\ast}t/\hbar}\right],\label{eq:Model_deviation_from_stationary_solution}
\end{align}
into Eqs.~(\ref{eq:Model_GP_equation_pm_1}) and (\ref{eq:Model_GP_equation_0}) and neglecting the higher-order terms of $u_m(x)$ and $v_m(x)$, we obtain the Bogoliubov equation
\begin{align}
&
\begin{bmatrix}
\bm{H}_1 & -\bm{H}_2 \\
\bm{H}_2^{\ast} & -\bm{H}_1^{\ast}
\end{bmatrix}
\begin{bmatrix}
\bm{u}(x) \\
\bm{v}(x)
\end{bmatrix}
=\epsilon
\begin{bmatrix}
\bm{u}(x) \\
\bm{v}(x)
\end{bmatrix}
,\label{eq:sp_Model_Bogoliubov_abbriviate_ver}\\
&\bm{H}_1\equiv 
\begin{bmatrix}
H_{11} & H_{12} & H_{13} \\
H^{\ast}_{12} & H_{22} & H_{23} \\
H^{\ast}_{13} & H^{\ast}_{23} & H_{33}
\end{bmatrix}
,\; \bm{H}_2\equiv
\begin{bmatrix}
H_{14} & H_{15} & H_{16} \\
H_{15} & H_{25} & H_{26} \\
H_{16} & H_{26} & H_{36}
\end{bmatrix}
,\label{eq:sp_Model_definition_of_H2}\\
&\bm{u}(x)=[u_{1}(x), u_{0}(x), u_{-1}(x)]^{\rm T},\label{eq:sp_Model_definiton_of_vec_u}\\
&\bm{v}(x)=[v_{1}(x), v_{0}(x), v_{-1}(x)]^{\rm T},\label{eq:sp_Model_definiton_of_vec_v}
\end{align}
where $\epsilon$ is the excitation energy. Here $H_{1, i j}$ and $H_{2, i j}$ are given by
\begin{align}
H_{11}&=-\frac{\hbar^2}{2M}\frac{d^2}{d x^2}-\mu+2(c_0+c_1)|\Psi_1(x)|^2\nonumber \\
&\quad +(c_0+c_1)|\Psi_0(x)|^2+(c_0-c_1)|\Psi_{-1}(x)|^2,\label{eq:sp_Model_H_11}\\
H_{12}&=(c_0+c_1)\Psi_0^{\ast}(x)\Psi_1(x)+2c_1\Psi_{-1}^{\ast}(x)\Psi_0(x),\label{eq:sp_Model_H_12}\\
H_{13}&=(c_0-c_1)\Psi_{-1}^{\ast}(x)\Psi_1(x),\label{eq:sp_Model_H_13}\\
H_{14}&=(c_0+c_1)\Psi_1(x)^2,\label{eq:sp_Model_H_14}\\
H_{15}&=(c_0+c_1)\Psi_1(x)\Psi_0(x),\label{eq:sp_Model_H_15}\\
H_{16}&=c_1\Psi_0(x)^2+(c_0-c_1)\Psi_1(x)\Psi_{-1}(x),\label{eq:sp_Model_H_16}\\
H_{22}&=-\frac{\hbar^2}{2M}\frac{d^2}{d x^2}-\mu+(c_0+c_1)|\Psi_1(x)|^2\nonumber\\
&\quad +2c_0|\Psi_0(x)|^2+(c_0+c_1)|\Psi_{-1}(x)|^2,\label{eq:sp_Model_H_22}\\
H_{23}&=(c_0+c_1)\Psi_{-1}^{\ast}(x)\Psi_0(x)+2c_1\Psi_0^{\ast}(x)\Psi_1(x),\label{eq:sp_Model_H_23}\\
H_{25}&=c_0\Psi_0(x)^2+2c_1\Psi_1(x)\Psi_{-1}(x),\label{eq:sp_Model_H_25}\\
H_{26}&=(c_0+c_1)\Psi_0(x)\Psi_{-1}(x),\label{eq:sp_Model_H_26}\\
H_{33}&=-\frac{\hbar^2}{2M}\frac{d^2}{d x^2}-\mu+(c_0-c_1)|\Psi_1(x)|^2\nonumber \\
&\quad +(c_0+c_1)|\Psi_0(x)|^2+2(c_0+c_1)|\Psi_{-1}(x)|^2,\label{eq:sp_Model_H_33}\\
H_{36}&=(c_0+c_1)\Psi_{-1}(x)^2.\label{eq:sp_Model_H_36}
\end{align}
We note that $\bm{H}_1^{\dagger}=\bm{H}_1$ and $\bm{H}_2^{\rm T}=\bm{H}_2$ hold, where $\dagger$ denotes the hermitian conjugate. The boundary conditions for $u_{m}(x)$ and $v_{m}(x)$ are given by
\begin{eqnarray}
u_{m}(x+L)&=&e^{+iM v L/\hbar}u_{m}(x),\label{eq:sp_Model_boundary_condition_for_u}\\
v_{m}(x+L)&=&e^{-iM v L/\hbar}v_{m}(x).\label{eq:sp_Model_boundary_condition_for_v}
\end{eqnarray}

Throughout this paper, we use the parameters $(L, c_1)=(96\xi_0, -0.005c_0)$, where $\xi_0\equiv \hbar/\sqrt{Mc_0n_0}$ is the healing length. These parameters correspond to the recent ring trap experiment \cite{Wright2013} and spin-1 ${}^{87}$Rb \cite{Kempen2002}. Our results presented below are valid for other parameter regions as long as $c_1<0$ and $c_0\gg |c_1|$.

\section{Results}\label{sec:Results}

\begin{figure}[t]
\centering
\includegraphics[width=8.5cm,clip]{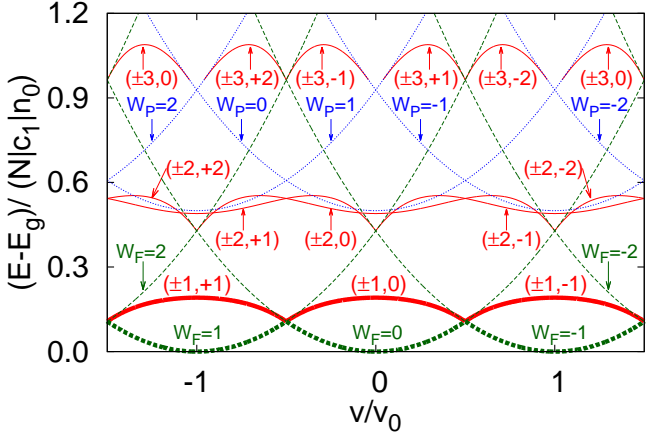}
\caption{(Color online) Velocity dependence of the energy per particle in the rotating frame for $L=96\xi_0$ and $c_1=-0.005c_0$, where $E_{\rm g}/N=(c_0+c_1)n_0/2$ denotes the ground-state energy. The solid red, dashed green, and dotted blue lines correspond to the TCPW states, FPW states, and PPW states, respectively. The thick (thin) lines represent the stable (unstable) branches. The integers in the figure represent the winding numbers $W_{\rm F}$, $W_{\rm P}$, and $(W, W_0)$.}
\label{fig:energy_diagram}
\vspace{-0.75em}
\end{figure}%

First, we present trivial plane-wave solutions of the GPE. It can be shown that ferromagnetic plane-wave (FPW)  and polar plane-wave (PPW) states
\begin{eqnarray}
\bm{\Psi}_{\rm F}(x)&=&\sqrt{n_0}e^{iM(v+W_{\rm F}v_0)x/\hbar}[1, 0, 0]^{\rm T},\label{eq:Results_FPW_solution}\\
\bm{\Psi}_{\rm P}(x)&=&\sqrt{n_0}e^{iM(v+W_{\rm P}v_0)x/\hbar}[0, 1, 0]^{\rm T},\label{eq:Results_PPW_solution}
\end{eqnarray}
are the stationary solutions of the GPE, where $W_{\rm F}, W_{\rm P}\in \mathbb{Z}$ are winding numbers and $n_0\equiv N/L$ is the mean-particle density. The velocity dependences of the energy for these states are shown in Fig.~\ref{fig:energy_diagram} \cite{Note_1}. We find that the ground state is the ferromagnetic state. The stability of these states can be determined by the Bogoliubov excitation spectra. The low-lying excitation of the ferromagnetic state is a magnon mode as long as $c_0\gg |c_1|$ and its expression is given by $\epsilon^{\rm M}_{\rm F}=-M v_0 |v+W_{\rm F}v_0|+Mv_0^2/2$. The low-lying magnon mode becomes negative, namely, Landau instability (LI) occurs for $|v+W_{\rm F}v_0|>v_0/2$. The low-lying excitations of the polar state are also magnon modes (doubly degenerated) and their expressions are given by $\epsilon^{\rm M}_{\rm P}=-Mv_0|v+W_{\rm P}v_0|+(Mv_0^2/2)\sqrt{1+4c_1n_0/Mv_0^2}$. Dynamical instability (DI) [${\rm Im}(\epsilon^{\rm M}_{\rm P})\not=0$] occurs in the polar state for $L/\xi_0>\pi\sqrt{c_0/|c_1|}$.

\begin{figure}[t]
\centering
\includegraphics[width=8.5cm,clip]{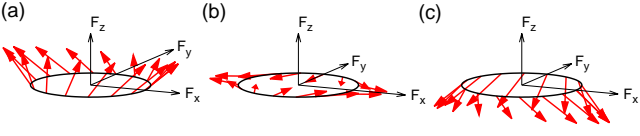}
\caption{(Color online) Spin textures for $W=-1$ branch at (a) $v=-0.3v_0$, (b) $v=0$, and (c) $v=0.3v_0$.}
\label{fig:schematic_texture}
\end{figure}

Next, we present a non-trivial solution of the GPE, which is expressed using the following ansatz:
\begin{eqnarray}
\Psi_m(x)=\sqrt{n_0}e^{iM v x/\hbar}e^{iM W_mv_0 x/\hbar}\phi_m,\label{eq:Results_ansatz_condensate_wave_function}
\end{eqnarray}
where, $W_m\in \mathbb{Z}$ is the winding number of the component $m$ and $\phi_m$ is a complex constant. We assume that all components of $\phi_m$ are non-zero. According to the ${\rm U(1)\times SO(3)}$ symmetry of the system and the requirement that the chemical potential be real valued, we can choose $\phi_1$ and $\phi_{0}$ to be real and positive and $\phi_{-1}$ to be real without loss of generality. The $\{\phi_m\}$ satisfy $\sum_m\phi_m^2=1$ due to the total particle number condition. When the winding numbers $\{W_m\}$ satisfy the relations $W\equiv W_0-W_1=W_{-1}-W_0\not=0$, the condensate wave function (\ref{eq:Results_ansatz_condensate_wave_function}) becomes a nontrivial solution of the GPE ($W=0$ states  correspond to trivial plane-wave states). The expressions of $\{\phi_m\}$ are given explicitly by
\begin{align}
\phi_{\pm 1}&=\left|\frac{W}{2}\pm\left(\frac{v}{v_0}+W_0\right)\right|\sqrt{\frac{M_z/N}{2W(v/v_0+W_0)}},\label{eq:Result_pm_1_component}\\
\phi_0&=\sqrt{1-\frac{M_z/N}{W(v/v_0+W_0)}\left[\frac{W^2}{4}+\left(\frac{v}{v_0}+W_0\right)^2\right]},\hspace{-0.5em}\label{eq:Results_0_component}\\
\frac{M_z}{N}&=\frac{2}{W}\left(\frac{v}{v_0}+W_0\right)\left[1-\frac{g(v, W, W_0)}{|c_1|n_0}\right],\label{eq:Results_expression_of_magnetization_of_z}
\end{align}
where $M_z\equiv \int^{L}_{0}dx F_z(x)=LF_z$ is the magnetization of the $z$ component ($F_z$ does not depend on $x$), $\phi_{-1}$ is positive due to Eq.~(\ref{eq:Results_existence_condition_2}), and $g(v, W, W_0)$ is a function defined for convenience as
\begin{eqnarray}
g(v, W, W_0)\equiv M\left[\frac{1}{4}W^2v_0^2-(v+W_0v_0)^2\right].\label{eq:Results_definition_of_function_g}
\end{eqnarray}
The parameter regions where the solution exists are given by $0\le \phi_0^2\le 1$:
\begin{align}
\hspace{-0.4em}\left|\frac{v}{v_0}+W_0\right|\le \frac{|W|}{2},\quad  \left(\frac{v}{v_0}+W_0\right)^2\ge \frac{W^2}{4}-\frac{|c_1|n_0}{Mv_0^2}.\label{eq:Results_existence_condition_2}
\end{align}
The explicit expressions of the physical quantities such as the chemical potential, total energy, total momentum of the rotating frame, and local magnetizations become
\begin{align}
\mu&=(c_0+c_1)n_0+\frac{1}{2}M(v+W_0v_0)^2+g(v, W, W_0),\label{eq:Results_expression_of_chemical_potential} \\
\frac{E}{N}&=\mu-\frac{1}{2}(c_0+c_1)n_0-\frac{g(v, W, W_0)^2}{2|c_1|n_0},\label{eq:Results_expression_of_total_energy}\\
P&\equiv -\frac{i\hbar}{2}\int^{L}_{0} dx\sum_m\left[\Psi^{\ast}_m(x)\frac{d}{d x}\Psi_m(x)-{\rm c. c.}\right] \notag \\
&=NM(v+W_0v_0)-MWv_0M_z,\label{eq:Results_expression_of_total_momentum}\\
F_x(x)&=\sqrt{2}n_0\phi_0(\phi_1+\phi_{-1})\cos\left(\frac{2\pi W}{L}x\right),\label{eq:Results_local_magnetization_x}\\
F_y(x)&=\sqrt{2}n_0\phi_0(\phi_1+\phi_{-1})\sin\left(\frac{2\pi W}{L}x\right).\label{eq:Results_local_magnetization_y}
\end{align}
From the expressions for the magnetization density (\ref{eq:Results_local_magnetization_x}) and (\ref{eq:Results_local_magnetization_y}), this solution [we call it the three-component plane-wave (TCPW) state] represents the spin texture. The spin rotates $|W|$ times around the ring. We plot the spin texture in Fig.~\ref{fig:schematic_texture}. This texture is similar to the polar core vortex (see Fig.~2 in Ref.~\cite{Isoshima2001}). We note that similar solutions for infinite systems were reported in Refs.~\cite{Cherng2008,Rodrigues2009,Tasgal2013,Tasgal2014a}.

The velocity dependences of the total energy for the TCPW states are shown by the solid red lines in Fig.~\ref{fig:energy_diagram}. At $v=-W_0v_0\pm v_0/2$, the TCPW states with $|W|=1$ branches appear [see Eq.~(\ref{eq:Results_existence_condition_2})]. These points are the same as the points at which LI occurs in the FPW branches due to the magnon mode, indicating that the instability of the magnon mode in the FPW state triggers the TCPW states, namely, the transition from the green line to the red line in Fig.~\ref{fig:energy_diagram}. For $|W|=3$ branches, the TCPW branches continuously connect the PPW branches. These bifurcations occur when the right-hand side of the second inequality in Eq.~(\ref{eq:Results_existence_condition_2}) becomes positive.

To investigate the stability of the TCPW states, we solve the Bogoliubov equation. From the boundary conditions (\ref{eq:sp_Model_boundary_condition_for_u}) and (\ref{eq:sp_Model_boundary_condition_for_v}), $\bm{u}(x)$ and $\bm{v}(x)$ can be expanded as a series of plane waves
\begin{align}
\bm{u}(x)&=e^{+iM(v+W_0v_0-Wv_0f_z)x/\hbar}\sum_n\frac{e^{i k_n x}}{\sqrt{L}}\bm{u}(k_n),\label{eq:sp_expansion_of_u}\\
\bm{v}(x)&=e^{-iM(v+W_0v_0-Wv_0f_z)x/\hbar}\sum_n\frac{e^{i k_n x}}{\sqrt{L}}\bm{v}(k_n),\label{eq:sp_expansion_of_v}
\end{align}
where $k_n\equiv 2\pi n/L$ is the wave number of the excitation and $n\in\mathbb{Z}$. Substituting Eqs.~(\ref{eq:sp_expansion_of_u}) and (\ref{eq:sp_expansion_of_v}) into Eq.~(\ref{eq:sp_Model_Bogoliubov_abbriviate_ver}) and using the wave function of the TCPW state (\ref{eq:Results_ansatz_condensate_wave_function}) and the translational symmetry of the system, we obtain the block-diagonalized Bogoliubov equation for $k$ space:
\begin{eqnarray}
\begin{bmatrix}
\bm{H}_1^+(k_n) & -\bm{H}_2(k_n) \\
\bm{H}_2(k_n) & -\bm{H}_1^{-}(k_n)
\end{bmatrix}
\begin{bmatrix}
\bm{u}(k_n) \\
\bm{v}(k_n)
\end{bmatrix}
=\epsilon_n
\begin{bmatrix}
\bm{u}(k_n) \\
\bm{v}(k_n)
\end{bmatrix}
,\label{eq:sp_Bogoliubov_equation_for_k_space}\\
\bm{H}_1^{\pm}(k_n)\equiv 
\begin{bmatrix}
H_{11}^{\pm}(k_n) & H_{12}(k_n) & H_{13}(k_n) \\
H_{12}(k_n) & H_{22}^{\pm}(k_n) & H_{23}(k_n) \\
H_{13}(k_n) & H_{23}(k_n) & H_{33}^{\pm}(k_n)
\end{bmatrix}
,\label{eq:sp_definition_of_H1_k}\\
\bm{H}_2(k_n)\equiv 
\begin{bmatrix}
H_{14}(k_n) & H_{15}(k_n) & H_{16}(k_n) \\
H_{15}(k_n) & H_{25}(k_n) & H_{26}(k_n) \\
H_{16}(k_n) & H_{26}(k_n) & H_{36}(k_n)
\end{bmatrix}
,\label{eq:sp_definition_of_H2_k}
\end{eqnarray}
where $H_{1, i j}^{\pm}(k_n)$ and $H_{2, i j}(k_n)$ are given by
\begin{align}
H_{11}^{\pm}(k_n)&=\epsilon_{k_n}^0\pm \hbar k_n(v+W_0v_0-Wv_0)\nonumber \\
&\quad +M\left(\frac{W}{2}v_0-W_0v_0-v\right)^2\nonumber \\
&\quad \quad +(c_0+c_1)n_0\phi_1^2-2c_1n_0\phi_{-1}^2,\label{eq:sp_Results_Bogoliubov_matrix_element_11}\\
H_{12}(k_n)&=(c_0+c_1)n_0\phi_1\phi_0+2c_1n_0\phi_0\phi_{-1},\label{eq:sp_Results_Bogoliubov_matrix_element_12}\\
H_{13}(k_n)&=(c_0-c_1)n_0\phi_1\phi_{-1},\label{eq:sp_Results_Bogoliubov_matrix_element_13}\\
H_{14}(k_n)&=(c_0+c_1)n_0\phi_1^2,\label{eq:sp_Results_Bogoliubov_matrix_element_14}\\
H_{15}(k_n)&=(c_0+c_1)n_0\phi_1\phi_0,\label{eq:sp_Results_Bogoliubov_matrix_element_15}\\
H_{16}(k_n)&=c_1n_0\phi_0^2+(c_0-c_1)n_0\phi_1\phi_{-1},\label{eq:Results_Bogoliubov_matrix_element_16}\\
H_{22}^{\pm}(k_n)&=\epsilon_{k_n}^0\pm \hbar k_n(v+W_0v_0)+M(v+W_0v_0)^2\nonumber \\
&\quad -\frac{1}{4}M(Wv_0)^2+(c_0-c_1)n_0\phi_0^2,\label{eq:sp_Results_Bogoliubov_matrix_element_22} \\
H_{23}(k_n)&=(c_0+c_1)n_0\phi_0\phi_{-1}+2c_1n_0\phi_1\phi_0,\label{eq:sp_Results_Bogoliubov_matrix_element_23}\\
H_{25}(k_n)&=c_0n_0\phi_0^2+2c_1n_0\phi_1\phi_{-1},\label{eq:sp_Results_Bogoliubov_matrix_element_25}\\
H_{26}(k_n)&=(c_0+c_1)n_0\phi_0\phi_{-1},\label{eq:sp_Results_Bogoliubov_matrix_element_26}\\
H_{33}^{\pm}(k_n)&=\epsilon_{k_n}^0\pm \hbar k_n(v+W_0v_0+Wv_0)\nonumber \\
&\quad +M\left(\frac{W}{2}v_0+W_0v_0+v\right)^2\nonumber \\
&\quad \quad -2c_1n_0\phi_1^2+(c_0+c_1)n_0\phi_{-1}^2,\label{eq:sp_Results_Bogoliubov_matrix_element_33} \\
H_{36}(k_n)&=(c_0+c_1)n_0\phi_{-1}^2,\label{eq:sp_Results_Bogoliubov_matrix_element_36}\\
\epsilon_{k_n}^0&\equiv \frac{\hbar^2k_n^2}{2M}.\label{eq:sp_free_particle_dispersion}
\end{align}
We note that $\bm{H}_1^{\pm}(k_n)$ and $\bm{H}_2(k_n)$ are the real and symmetric matrices.

\begin{figure}[t]
\centering
\includegraphics[width=8.5cm,clip]{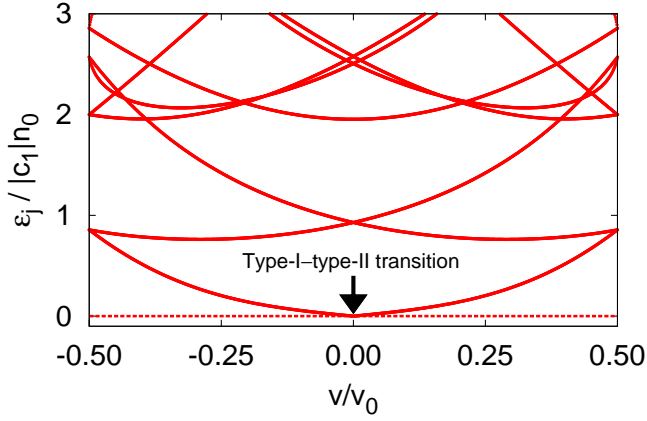}
\caption{(Color online) Velocity dependence of the excitation energy $\epsilon_j$ for $(L, c_1, W, W_0)=(96\xi_0, -0.005c_0,-1, 0)$. The dashed line denotes the zero modes and the arrow in the figure indicates the type-I$-$type-II transition point.}
\label{fig:excitation_L96c1m0005Wm1W00}
\end{figure}

\begin{figure}[t]
\includegraphics[width=8.5cm,clip]{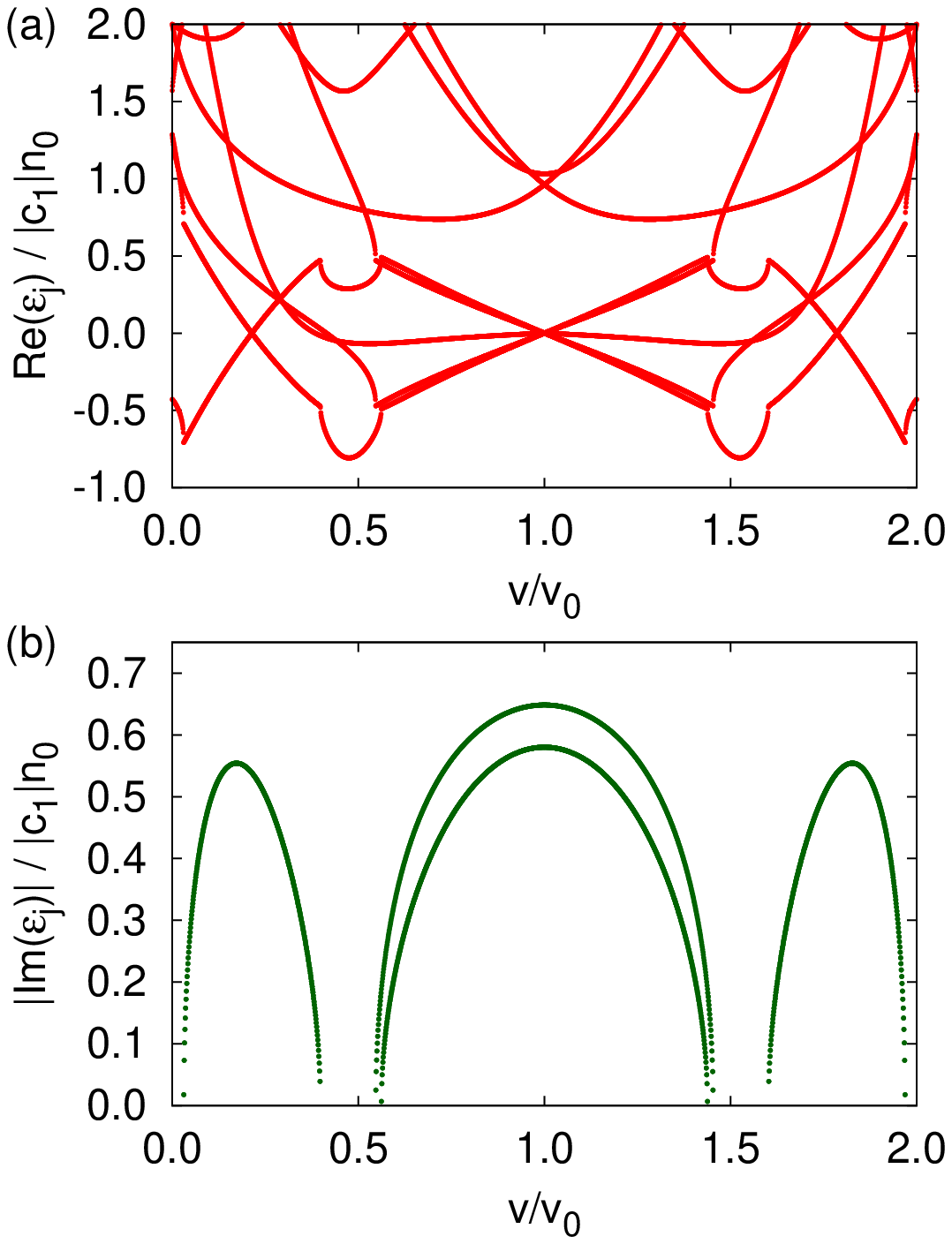}
\caption{(Color online) Velocity dependence of the (a) real part and (b) imaginary part of the excitation spectra for $(L, c_1, W, W_0)=(96\xi_0, -0.005c_0, -2, -1)$.}
\label{fig:sp_excitation_Wm2W0m1}
\end{figure}

\begin{figure}[t]
\includegraphics[width=8.5cm,clip]{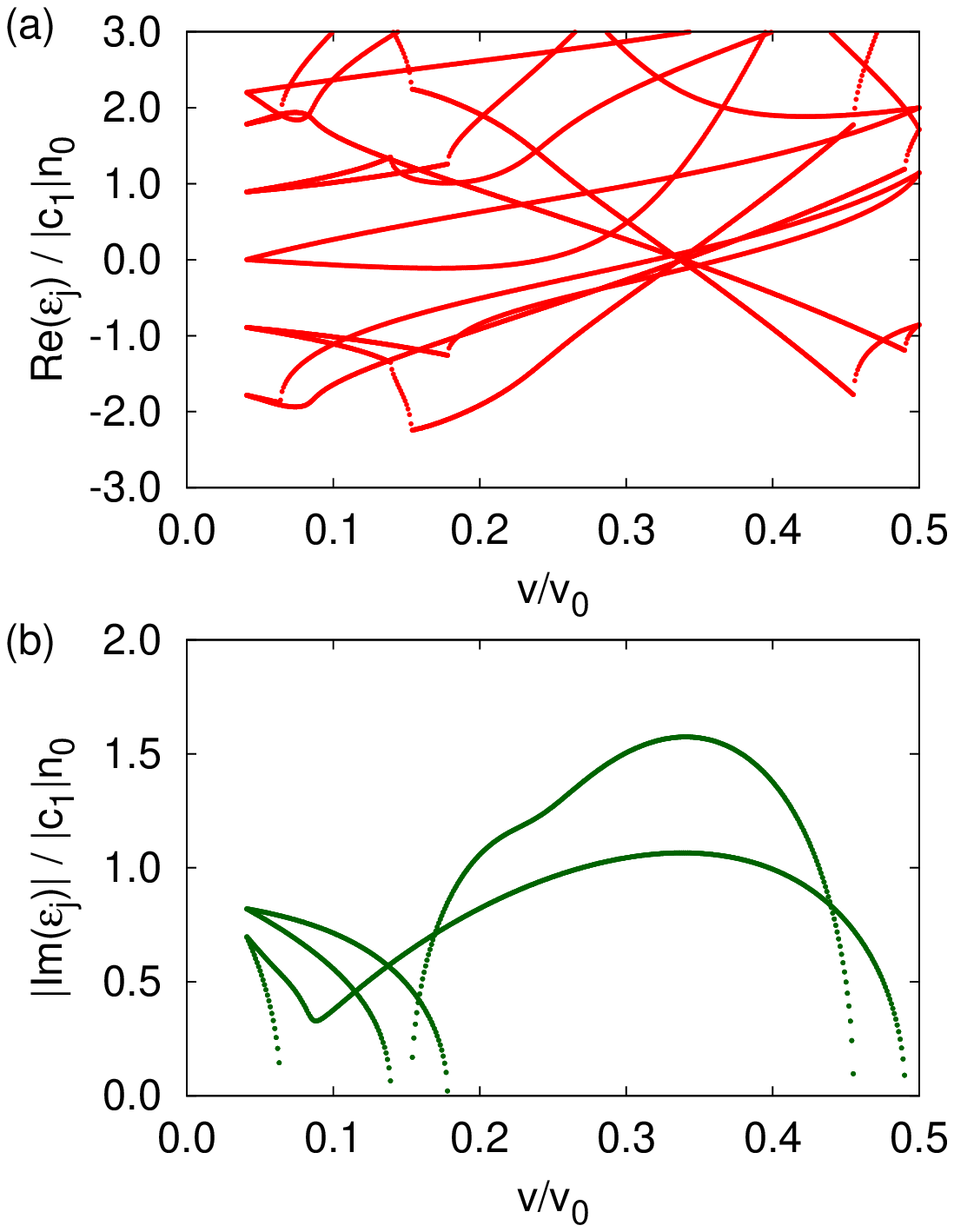}
\caption{(Color online) Velocity dependence of the (a) real part and (b) imaginary part of the excitation spectra for $(L, c_1, W, W_0)=(96\xi_0, -0.005c_0, -3, +1)$.}
\label{fig:sp_excitation_Wm3W01}
\end{figure}

The velocity dependence of the excited energy is shown in Figs.~\ref{fig:excitation_L96c1m0005Wm1W00}, \ref{fig:sp_excitation_Wm2W0m1}, and \ref{fig:sp_excitation_Wm3W01}. We find that the branch for $|W|=1$ is a metastable state because the energy is higher than the FPW branches and all excitation energies $\epsilon_j$ are real and positive, where $\epsilon_j$ is the $j$th excited energy and corresponding eigenvector is denoted by $\bm{x}_j\equiv [\bm{u}_j(x)^{\rm T}, \bm{v}_j(x)^{\rm T}]^{\rm T}$. We also find that LI or DI occurs in the branches for $|W|>1$ (see Figs.~\ref{fig:sp_excitation_Wm2W0m1} and \ref{fig:sp_excitation_Wm3W01}). The existence of the metastable spin texture state at the zero magnetic field is different point from the previous works \cite{Cherng2008,Tasgal2013,Tasgal2014a}. This is due to finite-size effects. Because the instability in infinite systems occurs at the long-wavelength limit, this instability is suppressed  in the finite-size systems.

We now discuss the NGMs of the TCPW states. The Hamiltonian of the system has ${\rm U(1)\times SO(3)}$ internal symmetry  \cite{Note_translation}; however, the TCPW state breaks this symmetry. Therefore, it can be expected that the system has a number of NGMs. 

First, we determine the number of the NGMs. According to recent developments of the theory of the NGMs \cite{Watanabe_Brauner2011,Watanabe_Murayama2012,Hidaka2013}, the NGMs can be classified as two different types: type I (unpaired) and type II (paired). The type-I NGM describes the mode related to one generator of the broken symmetry $Q_i$. On the other hand, the type-II NGM describes one mode related to a canonical pair of the two broken generators, which is analogous to the quantum mechanics; two variables $\hat{x}$ and $\hat{p}$ describe one degrees of freedom because these are the canonical pair; $[\hat{x}, \hat{p}]=i\hbar$. The total number of the NGMs can be obtained by using the Watanabe-Brauner (WB) matrix \cite{Watanabe_Brauner2011,Watanabe_Murayama2012,Hidaka2013}
\begin{eqnarray} 
\rho_{i j}\equiv -\frac{i}{L}\int^L_0dx\bm{\Psi}(x)^{\dagger}[Q_i, Q_j]\bm{\Psi}(x).\label{eq:Results_definition_WB_matrix}
\end{eqnarray}
In the present case, $\{Q_i\}$ is given by the spin matrices $f_x$, $f_y$, and $f_z$ and the $3\times 3$ unit matrix $I$. The total number of NGMs is given by $n_{{\rm NGM}}\equiv n_{\rm BG}-{\rm rank}(\rho)/2$, where $n_{\rm BG}\;(=4 \text{ in the present case})$ is  the number of the broken generators. Here $\rm{rank}(\rho)/2$ is the number of type-II NGMs.

In the present case, $\rho$ reduces to
\begin{align}
\rho=\frac{1}{L}
\begin{bmatrix}
0 & +M_z & 0 & 0 \\
-M_z & 0 & 0 & 0 \\
0 & 0 & 0 & 0 \\
0 & 0 & 0 & 0
\end{bmatrix}
,\label{eq:WB_matrix_present_case}
\end{align}
where we used $M_x=M_y=0$ for the TCPW state. Therefore, we obtain the number of the NGM
\begin{align}
n_{\rm NGM}=4-\frac{1}{2}{\rm rank}(\rho)=
\begin{cases}
3 \text{ for } M_z\not=0\\
4 \text{ for } M_z=0
\end{cases}
.\label{eq:Results_number_of_type_II}
\end{align}
This result shows that the number of NGMs changes at $M_z=0$ $(v=-W_0v_0$ \cite{Note_2}). Although the number of NGMs changes if the broken continuous symmetry changes, the broken continuous symmetry does not change at $M_z=0$ in the present case. This is called a type-I$-$type-II transition \cite{Takahashi2014a}. We will see what happens at $M_z=0$ by calculating the wave functions below.

Following Ref.~\cite{Takahashi2014a}, the zero-energy eigenstates of the Bogoliubov equation (zero modes) originating from the spontaneous symmetry breaking are given by using the broken generators and the order parameter: 
\begin{widetext}
\begin{align}
\bm{x}_{\rm B}&=
\begin{bmatrix}
\bm{\Psi}(x) \\
\bm{\Psi}^{\ast}(x)
\end{bmatrix}
=\sqrt{n_0}
\begin{bmatrix}
e^{+iM(v+W_0v_0-Wv_0f_z)x/\hbar}\bm{\phi}\\
e^{-iM(v+W_0v_0-Wv_0f_z)x/\hbar}\bm{\phi}
\end{bmatrix}
,\label{eq:explicit_sp_expression_B}\\
\bm{x}_z&=
\begin{bmatrix}
f_z\bm{\Psi}(x) \\
f_z^{\ast}\bm{\Psi}^{\ast}(x)
\end{bmatrix}
=\sqrt{n_0}
\begin{bmatrix}
e^{+iM(v+W_0v_0-Wv_0f_z)x/\hbar}f_z\bm{\phi} \\
e^{-iM(v+W_0v_0-Wv_0f_z)x/\hbar}f_z\bm{\phi}
\end{bmatrix}
,\label{eq:explicit_sp_expression_z}\\
\bm{x}_x&=
\begin{bmatrix}
f_x\bm{\Psi}(x)\\
f_x^{\ast}\bm{\Psi}^{\ast}(x)
\end{bmatrix}
\nonumber\\
&=\phantom{i}\sqrt{\frac{n_0}{2}}\left\{+e^{+iMWv_0x/\hbar}
\begin{bmatrix}
e^{+iM(v+W_0v_0-Wv_0f_z)x/\hbar}\bm{\phi}_+\\
e^{-iM(v+W_0v_0-Wv_0f_z)x/\hbar}\bm{\phi}_-
\end{bmatrix}
+e^{-iMWv_0x/\hbar}
\begin{bmatrix}
e^{+iM(v+W_0v_0-Wv_0f_z)x/\hbar}\bm{\phi}_-\\
e^{-iM(v+W_0v_0-Wv_0f_z)x/\hbar}\bm{\phi}_+
\end{bmatrix}
\right\}
,\label{eq:sp_explicit_expression_x}\\
\bm{x}_y&=
\begin{bmatrix}
f_y\bm{\Psi}(x) \\
f_y^{\ast}\bm{\Psi}^{\ast}(x)
\end{bmatrix}
\nonumber \\
&=i\sqrt{\frac{n_0}{2}}
\left\{-e^{+iMWv_0x/\hbar}
\begin{bmatrix}
e^{+iM(v+W_0v_0-Wv_0f_z)x/\hbar}\bm{\phi}_+\\
e^{-iM(v+W_0v_0-Wv_0f_z)x/\hbar}\bm{\phi}_-
\end{bmatrix}
+e^{-iMWv_0x/\hbar}
\begin{bmatrix}
e^{+iM(v+W_0v_0-Wv_0f_z)x/\hbar}\bm{\phi}_-\\
e^{-iM(v+W_0v_0-Wv_0f_z)x/\hbar}\bm{\phi}_+
\end{bmatrix}
\right\}
,\label{eq:sp_explicit_expression_y}\\
\bm{\phi}&\equiv [\phi_1, \phi_0, \phi_{-1}]^{\rm T},\label{eq:Results_definition_of_vector_form_phi}\\
\bm{\phi}_+&\equiv [\phi_0, \phi_{-1},0]^{\rm T},\label{eq:Results_definition_phi_+}\\
\bm{\phi}_-&\equiv [0, \phi_1,\phi_0]^{\rm T},\label{eq:Results_definition_phi_-}
\end{align}
\end{widetext}
which represent the global phase transformation and the spin rotation around the $x$, $y$, and $z$ axes, respectively. We can show that $\bm{x}_x$ and $\bm{x}_y$ are not independent for $M_z\not=0$. To see this, we calculate the following inner product \cite{Pethick2002,Kawaguchi2012,Takahashi2014a}: 
\begin{eqnarray}
(\bm{x}_i, \bm{x}_j)&\equiv& \int^L_0dx\bm{x}_i^{\dagger}\sigma_3\bm{x}_j,\label{eq:Results_definition_of_inner_product}\\
\sigma_3&\equiv &
\begin{bmatrix}
+I & 0 \\
0 & -I
\end{bmatrix}
.\label{eq:Results_definition_of_sigma_3}
\end{eqnarray}
By direct calculations, we can show that $(\bm{x}_x, \bm{x}_y)=(\bm{x}_y, \bm{x}_x)^{\ast}=iM_z$ and otherwise zero. These results mean that $\bm{x}_{\rm B}$ and $\bm{x}_z$ are always orthogonal to all other zero　modes and hence they give type-I NGMs. On the other hand, $\bm{x}_x$ and $\bm{x}_y$ give one type-II NGM or two type-I NGMs. For $M_z\not=0$, we can construct the wave function with positive norm by taking the linear combination for $\bm{x}_x$ and $\bm{x}_y$: $\bm{x}_{\pm}\equiv (\bm{x}_x\pm i\bm{x}_y)/\sqrt{2}$ \cite{Takahashi2014a}:
\begin{widetext}
\begin{align}
\bm{x}_+&=\frac{1}{\sqrt{2}}
\begin{bmatrix}
(f_x+if_y)\bm{\Psi}(x) \\
(f_x^{\ast}+if_y^{\ast})\bm{\Psi}^{\ast}(x)
\end{bmatrix}
=\sqrt{n_0}e^{+iMWv_0x/\hbar}
\begin{bmatrix}
e^{+iM(v+W_0v_0-Wv_0f_z)x/\hbar}\bm{\phi}_+\\
e^{-iM(v+W_0v_0-Wv_0f_z)x/\hbar}\bm{\phi}_-
\end{bmatrix}
,\label{eq:sp_explicit_expression_+}
\\
\bm{x}_-&=\frac{1}{\sqrt{2}}
\begin{bmatrix}
(f_x-if_y)\bm{\Psi}(x) \\
(f_x^{\ast}-if_y^{\ast})\bm{\Psi}^{\ast}(x)
\end{bmatrix}
=\sqrt{n_0}e^{-iMWv_0x/\hbar}
\begin{bmatrix}
e^{+iM(v+W_0v_0-Wv_0f_z)x/\hbar}\bm{\phi}_-\\
e^{-iM(v+W_0v_0-Wv_0f_z)x/\hbar}\bm{\phi}_+
\end{bmatrix}
.\label{eq:sp_explicit_expression_-}
\end{align}
\end{widetext}
The norm of $\bm{x}_{\pm}$ is given by $(\bm{x}_{\pm}, \bm{x}_{\pm})=\mp M_z$. Therefore, for $M_z>0\;(<0)$, only $\bm{x}_-\;(\bm{x}_+)$ is the physically meaningful mode and it gives the type-II NGM. On the other hand, for $M_z=0$, the following relations hold: $(\bm{x}_{\pm}, \bm{x}_{\mp})=(\bm{x}_{\pm}, \bm{x}_{\pm})=0$. Therefore, $\bm{x}_{+}$ and $\bm{x}_-$ are the type-I NGMs for $M_z=0$. 

\begin{figure}[t]
\centering
\includegraphics[width=8.0cm,clip]{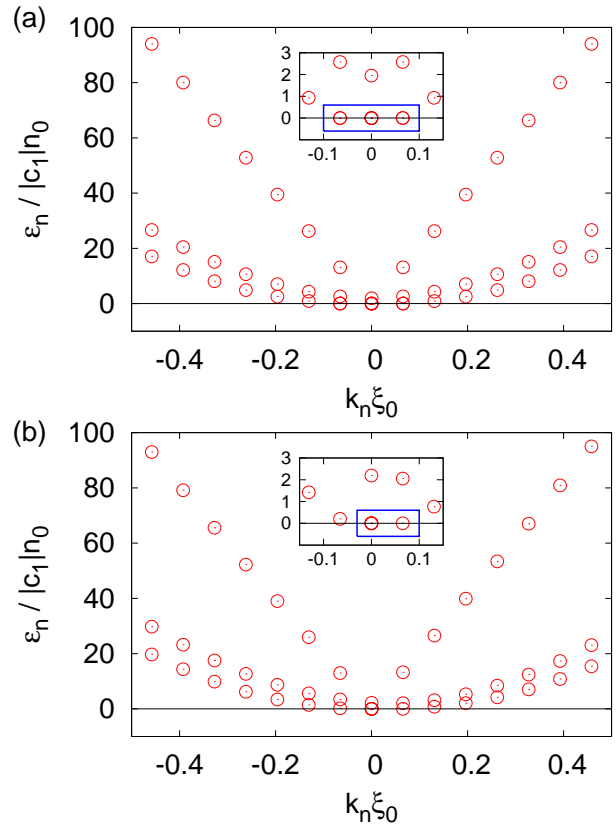}
\caption{(Color online) Excitation energy as a function of $k_n$ for $(L, c_1, W, W_0)=(96\xi_0, -0.005c_0,-1,0)$ at (a) $v=0$ (the type-I$-$type-II transition point) and (b) $v=-0.25v_0$. The insets show the magnified view around $k_n=0$. The points enclosed by blue boxes are zero-modes.}
\label{fig:sp_dispersion_relation}
\end{figure}

We also determine the wave number of the zero modes. By comparing the expressions of the zero modes with the expansion of the wave function for the excited states (\ref{eq:sp_expansion_of_u}) and (\ref{eq:sp_expansion_of_v}), we obtain the wave number of zero modes. We find that $\bm{x}_{\rm B}$ and $\bm{x}_z$ have zero wave number and $\bm{x}_{+}\;(\bm{x}_-)$ has the wave number $k_{+W}\;(k_{-W})$. These results can be seen in Fig.~\ref{fig:sp_dispersion_relation}. We summarize our results for the NGMs in Table~\ref{tab:summary_of_NGM}.

Here we discuss the discrete symmetry of the system. At the type-I$-$type-II transition point, the order parameter is invariant under the $\pi$ rotation around the $z$ axis and time-reversal operation: $\bm{\Psi}(x)=e^{-if_z\pi}\mathcal{T}\bm{\Psi}(x)$, where $\mathcal{T}$ is the time-reversal operator. This is because the twisted boundary condition (\ref{eq:Model_twisted_periodic_boundary_condition}) reduces to the periodic boundary condition at the transition point. According to this symmetry, both $\bm{x}_+$ and $\bm{x}_-$ become physically meaningful modes because these modes have the opposite wave vector.

\begin{table}[t]
\centering
\caption{Summary of the NGMs for the TCPW states.}
\begin{tabular}{ccccc}\hline
& {\footnotesize Type-I ($k_n=0$)} & {\footnotesize Type-I ($k_n\not=0$)}　& {\footnotesize Type-II ($k_n\not=0$)} \\ \hline
{\footnotesize $M_z>0$}& \footnotesize{$\bm{x}_{\rm B}, \bm{x}_z$} &  & \footnotesize{$\bm{x}_-$} \\
{\footnotesize $M_z=0$}& \footnotesize{$\bm{x}_{\rm B}, \bm{x}_z$} & \footnotesize{$\bm{x}_+$, $\bm{x}_-$} &  \\
{\footnotesize $M_z<0$}& \footnotesize{$\bm{x}_{\rm B}, \bm{x}_z$} &  & \footnotesize{$\bm{x}_+$} \\ \hline
\end{tabular}
\label{tab:summary_of_NGM}
\end{table}

Finally, we discuss the applicability of the above results to actual experiments. The TCPW states have different winding numbers, namely, different angular momenta for each component. Such states can be prepared through two-photon Raman transitions with circularly polarized Laguerre-Gaussian and standard Gaussian beams \cite{Wright2009}. The type-I$-$type-II transition can be indirectly observed by utilizing the vanishment of the first excited energy (see Fig.~\ref{fig:excitation_L96c1m0005Wm1W00}) because the first excited state, which is a magnetic excitation, is converted to the zero mode at $v=-W_0v_0$, as described above. The magnon excited energy and its dispersion relation have been observed in a recent experiment using a magnon contrast interferometry technique \cite{Marti2014}. By applying this technique to ring trap experiments, the type-I$-$type-II transition can be observed. 

\section{Summary}\label{sec:Summary}

We investigated the metastability of the spin textures and excitations of ferromagnetic spin-1 BECs confined in a rotating ring trap using mean-field theory. We found analytical solutions of the GPE (TCPW solutions) that exhibit spin textures analogous to polar-core vortices. By numerically solving the Bogoliubov equation, it was shown how the TCPW states can become metastable states. On the basis of these analytical solutions, we determined the number of NGMs by using the WB matrix \cite{Watanabe_Brauner2011,Watanabe_Murayama2012,Hidaka2013}. We found that the number of the NGMs changes at $v=-W_0v_0$ without changing the continuous symmetry of the order parameter.

In future work we hope to study the stability of the TCPW states in the presence of an external potential that breaks translational symmetry. An additional goal would be to perform a many-body calculation for spinor Bose gases in a ring trap.

\begin{acknowledgments}
The author thanks D. A. Takahashi, H. Saito, and Y. Kato for fruitful discussions. This work was supported by a Grant-in-Aid for Scientific Research on Innovative Areas ``Fluctuation \& Structure" (Grant No. 25103007) from the Ministry of Education, Culture, Sports, Science, and Technology of Japan.
\end{acknowledgments}



\begin{thebibliography}{99}
\bibitem{Ryu2007}
C. Ryu, M. F. Andersen, P. Clad\'{e}, V. Natarajan, K. Helmerson, and W. D. Phillips, Phys. Rev. Lett. {\bf 99}, 260401 (2007).
\bibitem{Ramanathan2011}
A. Ramanathan, K. C. Wright, S. R. Muniz, M. Zelan, W. T. Hill, C. J. Lobb, K. Helmerson, W. D. Phillips, and G. K. Campbell, Phys. Rev. Lett. {\bf 106}, 130401 (2011).
\bibitem{Moulder2012}
S. Moulder, S. Beattie, R. P. Smith, N. Tammuz, and Z. Hadzibabic, Phys. Rev. A {\bf 86}, 013629 (2012).
\bibitem{Beattie2013}
S. Beattie, S. Moulder, R. J. Fletcher, and Z. Hadzibabic, Phys. Rev. Lett. {\bf 110}, 025301 (2013).
\bibitem{Neely2013}
T. W. Neely, A. S. Bradley, E. C. Samson, S. J. Rooney, E. M. Wright, K. J. H. Law, R. Carretero-Gonz\'{a}lez, P. G. Kevrekidis, M. J. Davis, and B. P. Anderson, Phys. Rev. Lett. {\bf 111}, 235301 (2013).
\bibitem{Ryu2013}
C. Ryu, P. W. Blackburn, A. A. Blinova, and M. G. Boshier, Phys. Rev. Lett. {\bf 111}, 205301 (2013).
\bibitem{Wright2013}
K. C. Wright, R. B. Blakestad, C. J. Lobb, W. D. Phillips, and G. K. Campbell, Phys. Rev. Lett. {\bf 110}, 025302 (2013).
\bibitem{Wright2013_full}
K. C. Wright, R. B. Blakestad, C. J. Lobb, W. D. Phillips, and G. K. Campbell, Phys. Rev. A {\bf 88}, 063633 (2013).
\bibitem{Ryu2014}
C. Ryu, K. C. Henderson, and M. G. Boshier, New J. Phys. {\bf 16}, 013046 (2014).
\bibitem{Eckel2014}
S. Eckel, J. G. Lee, F. Jendrzejewski, N. Murray, C. W. Clark, C. J. Lobb, W. D. Phillips, M. Edwards, and G. K. Campbell, Nature (London) {\bf 506}, 200 (2014).
\bibitem{Jendrzejewski2014}
F. Jendrzejewski, S. Eckel, N. Murray, C. Lanier, M. Edwards, C. J. Lobb, and G. K. Campbell, Phys. Rev. Lett. {\bf 113}, 045305 (2014).
\bibitem{Eckel2014_2}
S. Eckel, F. Jendrzejewski, A. Kumar, C. J. Lobb, and G. K. Campbell, Phys. Rev. X {\bf 4}, 031052 (2014).
\bibitem{Corman2014}
L. Corman, L. Chomaz, T. Bienaim\'{e}, R. Desbuquois, C. Weitenberg, S. Nascimb\`{e}ne, J. Dalibard, and J. Beugnon, Phys. Rev. Lett. {\bf 113}, 135302 (2014).
\bibitem{Kawaguchi2012}
Y. Kawaguchi and M. Ueda, Phys. Rep. {\bf 520}, 253 (2012).
\bibitem{Stamper-Kurn2013}
D. M. Stamper-Kurn and M. Ueda, Rev. Mod. Phys. {\bf 85}, 1191 (2013).
\bibitem{Stamper-Kurn1998}
D. M. Stamper-Kurn, M. R. Andrews, A. P. Chikkatur, S. Inouye, H.-J. Miesner, J. Stenger, and W. Ketterle, Phys. Rev. Lett. {\bf 80}, 2027 (1998).
\bibitem{Stenger1998}
J. Stenger, S. Inouye, D. M. Stamper-Kurn, H. -J. Miesner, A. P. Chikkatur, and W. Ketterle, Nature (London) {\bf 396}, 345 (1998).
\bibitem{Leonhardt2000}
U. Leonhardt and G. E. Volovik, JETP. Lett. {\bf 72}, 46 (2000).
\bibitem{Khawaja2001}
U. Al Khawaja and H. Stoof, Nature (London) {\bf 411}, 918 (2001).
\bibitem{Leanhardt2003}
A. E. Leanhardt, Y. Shin, D. Kielpinski, D. E. Pritchard, and W. Ketterle, Phys. Rev. Lett. {\bf 90}, 140403 (2003).
\bibitem{Ruostekoski2003}
J. Ruostekoski and J. R. Anglin, Phys. Rev. Lett. {\bf 91}, 190402 (2003).
\bibitem{Kawaguchi2008}
Y. Kawaguchi, M. Nitta, and M. Ueda, Phys. Rev. Lett. {\bf 100}, 180403 (2008).
\bibitem{Kobayashi2009}
M. Kobayashi, Y. Kawaguchi, M. Nitta, and M. Ueda, Phys. Rev. Lett. {\bf 103}, 115301 (2009).
\bibitem{Leslie2009}
L. S. Leslie, A. Hansen, K. C. Wright, B. M. Deutsch, and N. P. Bigelow, Phys. Rev. Lett. {\bf 103}, 250401 (2009).
\bibitem{Wright2009}
K. C. Wright, L. S. Leslie, A. Hansen, and N. P. Bigelow, Phys. Rev. Lett. {\bf 102}, 030405 (2009).
\bibitem{Choi2012}
J. Y.  Choi, W. J. Kwon, and Y. I. Shin, Phys. Rev. Lett. {\bf 108}, 035301 (2012).
\bibitem{Choi2012_2}
J. Y. Choi, W. J. Kwon, M. Lee, H. Jeong, K. An, and Y. I. Shin, New. J. Phys. {\bf 14}, 053013 (2012).
\bibitem{Ray2014}
M. W. Ray, E. Ruokokoski, S. Kandel, M. M\"{o}tt\"{o}nen, and D. S. Hall, Nature (London) {\bf 505}, 657 (2014).
\bibitem{Sadler2006}
L. E. Sadler, J. M. Higbie, S. R. Leslie, M. Vengalattore, and D. M. Stamper-Kurn, Nature (London) {\bf 443}, 312 (2006).
\bibitem{Vengalattore2008}
M. Vengalattore, S. R. Leslie, J. Guzman, and D. M. Stamper-Kurn, Phys. Rev. Lett. {\bf 100}, 170403 (2008).
\bibitem{Cherng2008}
R. W. Cherng, V. Gritsev, D. M. Stamper-Kurn, and E. Demler, Phys. Rev. Lett. {\bf 100}, 180404 (2008).
\bibitem{Lamacraft2008}
A. Lamacraft, Phys. Rev. A {\bf 77}, 063622 (2008).
\bibitem{Vengalattore2010}
M. Vengalattore, J. Guzman, S. R. Leslie, F. Serwane, and D. M. Stamper-Kurn, Phys. Rev. A {\bf 81}, 053612 (2010).
\bibitem{Ohmi1998}
T. Ohmi and K. Machida, J. Phys. Soc. Jpn. {\bf 67}, 1822 (1998).
\bibitem{Ho1998}
T. L. Ho, Phys. Rev. Lett. {\bf 81}, 742 (1998).
\bibitem{Murata2007}
K. Murata, H. Saito, and M. Ueda, Phys. Rev. A {\bf 75}, 013607 (2007).
\bibitem{Uchino2010}
S. Uchino, M. Kobayashi, and M. Ueda, Phys. Rev. A {\bf 81}, 063632 (2010).
\bibitem{Symes2014}
L. M. Symes, D. Baillie, and P. B. Blakie, Phys. Rev. A {\bf 89}, 053628 (2014).
\bibitem{Marti2014}
G. E. Marti, A. MacRae, R. Olf, S. Lourette, F. Fang, and D. M. Stamper-Kurn, Phys. Rev. Lett. {\bf 113}, 155302 (2014).
\bibitem{Smyrnakis2009}
J. Smyrnakis, S. Bargi, G. M. Kavoulakis, M. Magiropoulos, K. K\"{a}rkk\"{a}inen, and S. M. Reimann, Phys. Rev. Lett. {\bf 103}, 100404 (2009).
\bibitem{Bargi2010}
S. Bargi, F. Malet, G. M. Kavoulakis, and S. M. Reimann, Phys. Rev. A {\bf 82}, 043631 (2010).
\bibitem{Anoshkin2013}
K. Anoshkin, Z. Wu, and E. Zaremba, Phys. Rev. A {\bf 88}, 013609 (2013).
\bibitem{Smyrnakis2013}
J. Smyrnakis, M. Magiropoulos, G. M.  Kavoulakis, and A. D. Jackson, Phys. Rev. A {\bf 87}, 013603 (2013).
\bibitem{Wu2013}
Z. Wu and E. Zaremba, Phys. Rev. A {\bf 88}, 063640 (2013).
\bibitem{Makela2013}
H. M\"akel\"a and E. Lundh, Phys. Rev. A {\bf 88}, 033622 (2013).
\bibitem{Yakimenko2013}
A. I. Yakimenko, K. O. Isaieva, S. I. Vilchinskii, and M. Weyrauch, Phys. Rev. A, {\bf 88}, 051602(R) (2013).
\bibitem{Mueller2002}
E. J. Mueller, Phys. Rev. A {\bf 66}, 063603 (2002).
\bibitem{Kanamoto2009}
R. Kanamoto, L. D. Carr, and M. Ueda, Phys. Rev. A {\bf 79}, 063616 (2009).
\bibitem{Takahashi2014a}
D. A. Takahashi and M. Nitta, arXiv:1404.7696, to appear in Ann. Phys. (doi : 10.1016/j.aop.2014.12.009).
\bibitem{Note_angle}
The relation between $x$ and the azimuthal angle $\theta$ is given by $x=R\theta$, where $R\;(=L/2\pi)$ is the radius of the ring. We assume that $R\gg r$, where $S\equiv \pi r^2$ and $r$ is the radius of the tube. This is the condition where the quasi-one-dimensional approximation is valid. In the current experiment, this condition is not satisfied: $R/r\sim 2$ \cite{Wright2013,Wright2013_full}.
\bibitem{Lieb2002_2}
E. H. Lieb, R. Seiringer, and J. Yngvason, Phys. Rev. B {\bf 66}, 134529 (2002).
\bibitem{Kunimi2014_1}
This boundary condition can be derived as follows: First, we consider the laboratory frame; in this frame, the periodic boundary condition is imposed on the condensate wave function $\bm{\Psi}_{\rm lab}(x_{\rm L}, t)$ because there is a requirement that $\bm{\Psi}_{\rm lab}(x_{\rm L}, t)$ be single valued, where $x_{\rm L}$ denotes the coordinate of the laboratory frame. The condensate wave function in the rotating frame with velocity $-v$ is given by $\bm{\Psi}(x, t)=e^{i(Mv^2t/2+M v x_{\rm L})/\hbar}\bm{\Psi}_{\rm lab}(x_{\rm L}, t)$, where $x\equiv x_{\rm L}+v t$. Therefore, $\bm{\Psi}(x, t)$ satisfies the twisted periodic boundary condition. See also, M. Kunimi and Y. Kato, arXiv:1407.7915.
\bibitem{Kempen2002}
E. G. M. van Kempen, S. J. J. M. F. Kokkelmans, D. J. Heinzen, and B. J. Verhaar, Phys. Rev. Lett. {\bf 88}, 093201 (2002).
\bibitem{Note_1}
Although other types of solution such as soliton states \cite{Zhang2012}, counterflow states \cite{Fujimoto2012_1}, and half-quantum vortex states \cite{Hoshi2008} can appear in Fig.~\ref{fig:energy_diagram}, we do not consider them in this paper.
\bibitem{Zhang2012}
Z. H. Zhang, C. Zhang, S. J. Yang, and S. Feng, J. Phys. {\bf 45}, 215302 (2012).
\bibitem{Fujimoto2012_1}
K. Fujimoto and M. Tsubota, Phys. Rev. A {\bf 85}, 033642 (2012).
\bibitem{Hoshi2008}
S. Hoshi and H. Saito, Phys. Rev. A {\bf 78}, 053618 (2008).
\bibitem{Isoshima2001}
T. Isoshima, K. Machida, and T. Ohmi, J. Phys. Soc. Jpn. {\bf 70}, 1604 (2001).
\bibitem{Rodrigues2009}
A. S. Rodrigues, P. G. Kevrekidis, R. Carretero-Gonz\'{a}lez, D. J. Frantzeskakis, P. Schmelcher, T. J. Alexander, and Yu. S. Kivshar, Phys. Rev. A, {\bf 79}, 043603 (2009).
\bibitem{Tasgal2013}
R. S. Tasgal and Y. B. Band, Phys. Rev. A {\bf 87}, 023626 (2013).
\bibitem{Tasgal2014a}
R. S. Tasgal and Y. B. Band, arXiv:1408.4594.
\bibitem{Note_translation}
The TCPW states spontaneously break the translational symmetry because they exhibit spin textures. However, an NGM originating from the translational symmetry does not exist because the translation of the TCPW can be written as $\bm{\Psi}(x+x_0)=e^{iM(v+W_0v_0)x_0/\hbar}e^{-i M W v_ 0f_z x_0/\hbar}\bm{\Psi}(x)$, where $x_0$ is a constant. This means that the translation can be represented by a combination of the ${\rm U(1)}$ gauge transformation and the spin rotation around $z$ axis. 
\bibitem{Watanabe_Brauner2011}
H. Watanabe and T. Brauner, Phys. Rev. D {\bf 84}, 125013 (2011).
\bibitem{Watanabe_Murayama2012}
H. Watanabe and H. Murayama, Phys. Rev. Lett. {\bf 108}, 251602 (2012).
\bibitem{Hidaka2013}
Y. Hidaka, Phys. Rev. Lett. {\bf 110}, 091601 (2013).
\bibitem{Note_2}
Although $M_z=0$ also holds at $(v/v_0+W_0)^2=W^2/4-|c_1|n_0/Mv_0^2$, we do not consider such a case because this point is a transition point from a TCPW state to a PPW state.
\bibitem{Pethick2002}
C. J. Pethick and H. Smith, {\it Bose-Einstein Condensation in Diluted Gases} (Cambridge University Press, Cambridge, 2002).

\end{thebibliography}


\end{document}